\documentclass[amsmath,aps,prl,twocolumn]{revtex4-1}
\usepackage{graphicx}
\usepackage{times}
\usepackage{bm}

\begin{document}
\title{Analysis of nonresonant effects in the two-photon spectroscopy of helium}
\author{T. Zalialiutdinov$^{1}$, A. Anikin$^{1}$, D. Solovyev$^1$}

\affiliation{ 
$^1$ Department of Physics, St. Petersburg State University, Petrodvorets, Oulianovskaya 1, 198504, St. Petersburg, Russia
}

\begin{abstract}
In the present paper, we study nonresonant corrections for experimental measurements of the transition frequencies in the helium atom. Having attracted more attention, such effects can make a significant contribution to experiments based on one- and two-photon atomic spectroscopy. The quantum interference effects in the measurements of $2^3S_1-n^3D_1$ ($ n=3,\,4,\,5 $) transition frequencies based on Doppler-free two-photon spectroscopy, are considered as a possible source of current discrepancy between the experimental and theoretical data. We demonstrate that line profile asymmetry caused by the quantum interference of fine sub-levels of the $ ^3D_{J} $ $ (J=1,2,3) $ state can reach tenths of a megahertz for different experimental conditions. Thus, previously unaccounted nonresonant corrections should be taken in next-generation experimental measurements of transitions frequencies in helium. However, they could not completely eliminate the current imbalance in the study of helium spectra and the question is still open.
\end{abstract}

\maketitle

Over the past two decades, the absolute measurement of transition frequencies in simple atomic systems like hydrogen and helium has been continuously improved with the aim of determining the fundamental physical constants, nuclear properties, testing quantum electrodynamics calculations. Recent precise measurements of transition frequencies in hydrogen have achieved a level where the quantum interference effects (QIE) between closest degenerate states have played a decisive role \cite{beyer,yost}. Significant progress in the spectroscopy of one-electron systems has spurred on the study of nonresonant (NR) corrections to the transition energies in many-electron systems \cite{Amaro-mH-2015,Brown-2013,MHH-2015,Sansonetti}. Although helium has been studied theoretically and experimentally for many years, accounting for the nonresonant effect \cite{Jent-Mohr} and the QIE as a part of them in spectroscopic measurements of the transition frequencies has not been considered until recently \cite{Amaro-mH-2015,MHH-2015}.

The energies of atomic levels in helium are conventionally expressed as the sum of nonrelativistic energies, lowest-order relativistic corrections, Lamb shift, etc., which includes quantum electrodynamics corrections (QED) and higher-order relativistic terms. Recent calculations of QED effects at the $ \alpha^7m $ level have improved theoretical predictions for the energies of helium atomic levels, leading to complete agreement with the measured $2^3S -2^3P$ transition frequency \cite{yerokhin2021}. However, as found in \cite{yerokhin2021}, such calculations do not eliminate the discrepancy between theoretical predictions and the experimental result for the $2^3S_{1} -3^3D_{1}$ transition discussed earlier in \cite{yerokhin2020}. 

This work is devoted to a theoretical analysis of the nonresonant effects arising from neighbouring fine structure sublevels, when measuring the energy of the $2^3S_{1} -3^3D_{1}$  transition, and is aimed at eliminating the emerged theoretical and experimental discrepancies by taking into account the QIE correction. In the experiment reported in \cite{Dorrer}, helium atoms in an atomic beam are prepared in metastable $ 2^3S_{1} $ state and then excited into the $ 3^3D_{1} $ state by absorbing two photons with equal frequencies, $ \omega_{1}=\omega_{2} $ having parallel polarizations $ \textbf{e}_1 $ and $ \textbf{e}_2 $, and propagating in opposite directions \cite{Schwob}. The detection of an excited fraction of $ 3^3D_{1} $ atoms is observed by fluorescence (i.e., decay into $ 2^3P $  states) with the emission of a photon with a frequency $ \omega_{3} $, polarization $ \textbf{e}_3^* $ towards direction $ \bm{\nu}_{3} $. In this case, there should be interference between sublevels of the fine structure $ 3^3D_{1} $, $ 3^3D_{2} $ and $ 3^3D_{3} $, which leads to an asymmetry of the line profile \cite{LSPS}. This experimental situation is similar to that previously considered in \cite{twophoton2021}, where NR corrections to the energies of $2s - ns/nd$ transitions in hydrogen ($n = 4,\; 6,\;8,\;12$ is the principal quantum number) with interference between $nd^{F =2}_{3/2}$ and $nd^{F =2}_{5/2}$ states were studied. Applying the QED theory of the spectral line profile, see \cite{Andr}, we describe the NR correction to the transition frequency $2^3S_{1} -3^3D_{1}$ for the experimental scheme given in \cite{Dorrer}. 

Following \cite{Dorrer} we are interested in the case when two incident photons with polarizations $ \textbf{e}_{1} $ and $ \textbf{e}_{2} $ propagating in opposite directions given by vectors $ \bm{\nu}_{1} $ and $ \bm{\nu}_{2} $, respectively, are absorbed from initially prepared $ 2^3S_{1} $ state into the intermediate state $3^3D_{J_{n}}$, thereat $ \omega_{1}+\omega_{2} = E_{3^3D_{J_{n}}}-E_{2^3S_{1}}$. Two-photon selection rules allow the transition to the different fine sublevels with $ J_{n}=1,\,2,\,3 $. Then the fluorescent signal in a direction $ \bm{\nu}_{3} $ of photons with frequency $ \omega_{3} $ and polarization $ \textbf{e}_{3}^* $ emitted in electric dipole transition $3^3D_{J_{n}}\rightarrow 2^3P_{J_{f}}+\gamma(\mathrm{E1})$ is observed. Performing summation over projections of angular momenta of final and intermediate states, averaging over projections of initial state, integrating over the frequency $ \omega_{3} $ the corresponding cross section of scattering process can be written as \cite{twophoton2021} 
\begin{widetext}
\begin{eqnarray}
\label{0}
\sigma(\omega_{1},\omega_{2})= \frac{\mathrm{const}}{2J_{i}+1}\sum\limits_{J_{f}}\sum\limits_{M_{J_{i}}M_{J_{f}}}
\left|
\sum\limits_{\substack{J_{n}M_{J_{n}}\\kJ_{k}M_{J_{k}}}}
(E_{2^3S_{1}}-E_{2^3P_{J_{f}}}+\omega_{1}+\omega_{2})^{3/2}\omega_{2}^{1/2}
\omega_{1}^{1/2}
\frac{\langle 2^3P_{J_{f}} | \textbf{e}_{3}^*\textbf{D} | 3^3D_{J_{n}} \rangle}{E_{3^3D_{J_{n}}}-E_{2^3S_{1}}-\omega_{1}-\omega_{2}}
\right.
\\\nonumber
\left.
\times
\frac{
	\langle 3^3D_{J_{n}} | \textbf{e}_{2}\textbf{D} | k^3P_{J_{k}} \rangle
	\langle k^3P_{J_{k}} | \textbf{e}_{1}\textbf{D}  | 2^3S_{1} \rangle
}
{
	E_{k^3P_{J_{k}}}-E_{2^3S_{1}}-\omega_{2}
}
+\mbox{5 permutations}
\right|^2
,
\end{eqnarray}
\end{widetext}
where $ \textbf{D}=e(\textbf{r}_1+\textbf{r}_2) $ is the electric dipole operator of two-electron system ($ e $ is the charge of electron) and '$\mathrm{const}$' is a some numerical factor which is unimportant for further calculations. Permutations in Eq. (\ref{0}) are understood as all possible rearrangement of indices 1, 2, 3 denoting the corresponding photons. Since the detector in the experiment \cite{Dorrer} do not resolve energies of fluorescent photon connected with the decay to different fine sublevels $J_{f}=0,\,1,\,2$ an additional summation over $ J_{f} $ appeared in Eq. (\ref{0}). Sum over $ k $ in Eq. (\ref{0}) implies summation over the entire spectrum of the Schr\"odinger equation for two-electron system including integration over the continuous spectrum. The calculation of total cross section for two- and three-photon scattering processes in helium with summation over the spectrum of fully correlated two-electron intermediate states was considered earlier in works \cite{DrakeHeliumTwo,SSSR-2016}. In Eq. (\ref{0}) we also neglect the contribution of higher-order multipoles, since their contribution to the NR effects is negligible at the current experimental accuracy \cite{LabKlim}.


The matrix elements of Eq. (\ref{0}) depend on the photon directions $ \bm{\nu}_1 $, $ \bm{\nu}_2 $ and $ \bm{\nu}_3 $ through the transversality condition of the polarization vectors. In the \cite{Dorrer} experiment, incident photons propagate in opposite directions with polarization vectors $ \textbf{e}_1 $ and $ \textbf{e}_2 $, and the outgoing photon has polarization $ \textbf{e}_3^* $, whose directions are considered fixed. Then, denoting the angles between any pair of two vectors as $ \theta_{ij} $ ($ i,j=1,\;2,\;3 $), we will be interested in the arsing angular correlations depending on these angles when determining the transition frequency from the cross section (\ref{0}).

Since the experiment \cite{Dorrer} is aimed at measuring the two-photon absorption transition $2^3S_1+2\gamma(\mathrm{E1})\rightarrow 3^3D_{1}$, we employ the resonant approximation \cite{Andr}. Within the framework of this simplification, the nonresonant terms describing transitions to other levels $3^3D_{J_{n}=2,\,3}$ in the scattering amplitude Eq. (\ref{0}) can be omitted with high accuracy, what is justified only if the corresponding NR corrections go beyond the accuracy of experiments \cite{S-2020-importance}. According to \cite{Dorrer}, the frequencies of two incident laser photons are equal
\begin{eqnarray}
\nonumber
\omega_{1}=\omega_{2}  = \omega = (E_{3^3D_{1}}-E_{2^3S_{1}})/2
,
\end{eqnarray}
and the detector of fluorescent signal is insensitive to the polarization $ \textbf{e}_{3}^* $ of the outgoing photon. Then, taking into account that $ \omega_{1}+\omega_{2}=E_{3^3D_{1}}-E_{2^3S_{1}} $ and summing over the polarization $ \textbf{e}_3^* $
with the use of equation \cite{LabKlim}
\begin{eqnarray}
\label{pol}
\sum\limits_{\textbf{e}}(\textbf{e}^*\textbf{D})(\textbf{e} \textbf{D})=(\bm{\nu}\times \textbf{D})(\bm{\nu}\times \textbf{D})
\end{eqnarray}
the cross section in the resonant approximation can be written in the form:
\begin{widetext}
\begin{eqnarray}
\label{1}
\sigma(\omega)= \frac{\mathrm{const}}{2J_{f}+1}
\sum\limits_{J_{f}}\sum\limits_{M_{J_{i}}M_{J_{f}}}
\left|
\sum\limits_{\substack{J_{n}M_{J_{n}}\\kJ_{k}M_{J_{k}}}}
(E_{3^3D_{1}}-E_{2^3P_{J_{f}}})^{3/2}
(E_{3^3D_{1}}-E_{2^3S_{1}}-\omega)^{1/2}
\omega^{1/2}
\right.
\\\nonumber
\times
\frac{\langle 2^3P_{J_{f}} | \bm{\nu}_{3}\times\textbf{D} | 3^3D_{J_{n}} \rangle}{E_{3^3D_{J_{n}}}-E_{2^3S_{1}}-2\omega-\frac{\mathrm{i}}{2}\Gamma_{3^3D_{J_{n}}}}
\left\lbrace
\frac{
	\langle 3^3D_{J_{n}} | \textbf{e}_{2}\textbf{D} | k^3P_{J_{k}} \rangle
	\langle k^3P_{J_{k}} | \textbf{e}_{1}\textbf{D}  | 2^3S_{1} \rangle
}
{
	E_{k^3P_{J_{k}}}-E_{2^3S_{1}}-\omega
}
\left.
+
\frac{
	\langle 3^3D_{J_{n}} | \textbf{e}_{1}\textbf{D}| k^3P_{J_{k}} \rangle
	\langle k^3P_{J_{k}} | \textbf{e}_{2}\textbf{D}| 2^3S_{1} \rangle
}
{
	E_{k^3P_{J_{k}}}-E_{3^3D_{J_{n}}}+\omega
}
\right\rbrace
\right|^2
,
\end{eqnarray}
\end{widetext}
where the regularization procedure established in \cite{Low} and assembled latter in \cite{Andr, ZSLP-report} was applied. This regularization procedure corresponds to the summation of an infinite number of one-loop self-energy insertions into the electron propagator. Within the framework of resonant approximation, this leads to the appearance of gauge invariant level widths $\Gamma_{3^3D_{J_{n}}}$ in the singular denominator Eq. (\ref{1}), which, in turn, leads to the formation of an absorption line profile.

To introduce NR corrections to the considered resonant transition $3^3D_{1}$ the remaining terms with $ J_{n}=2 $ and $J_{n}=3$ in the scattering amplitude Eq. (\ref{1}) should be considered. It can be found that the neighboring fine states presented in the sum over $ J_{n} $ provide the leading NR contribution \cite{Jent-Mohr}, which is now known as the effect of quantum interference. Then, following the methodology given in \cite{twophoton2021}, denoting the intervals $ \Delta_{12} = E_{3^3D_{1}}-E_{3^3D_{2}} $ and $ \Delta_{13} = E_{3^3D_{1}}-E_{3^3D_{3}}$, summing over the projections of final state and average over the projections of initial state, we arrive at
\begin{eqnarray}
\label{11}
\sigma(\omega)=\mathrm{const}
\left[
\frac{f_{\mathrm{res}}^{3^3D_{1}}}{(\omega_{0}-2\omega)^2+\frac{\Gamma_{3^3D_{1}}^2}{4}}
\right.
\\\nonumber
+2\mathrm{Re}\frac{f_{\mathrm{nr}}^{3^3D_{1(2)}}}{(\omega_{0}-2\omega-\frac{\mathrm{i}\Gamma_{3^3D_{1}}}{2})(\omega_{0}+\Delta_{12}-2\omega)}
\\\nonumber
\left.
+2\mathrm{Re}\frac{f_{\mathrm{nr}}^{3^3D_{1(3)}}}{(\omega_{0}-2\omega-\frac{\mathrm{i}\Gamma_{3^3D_{1}}}{2})(\omega_{0}+\Delta_{13}-2\omega)}
\right]
,
\end{eqnarray}
where we introduced following notations
\begin{eqnarray}
\label{fres}
f^{3^3D_{1}}_{\mathrm{res}}=\sum\limits_{M_{J_i}M_{J_f}}|T_{1}(\omega_{0}/2)|^2,
\end{eqnarray}
\begin{eqnarray}
\label{fnr12}
f^{3^3D_{1(2)}}_{\mathrm{nr}} =\sum\limits_{M_{J_i}M_{J_f}}
T_{1}(\omega_{0}/2)
T^*_{2}(\omega_{0}/2),
\end{eqnarray}
\begin{eqnarray}
\label{fnr13}
f^{3^3D_{1(3)}}_{\mathrm{nr}} =\sum\limits_{M_{J_i}M_{J_f}}
T_{1}(\omega_{0}/2)
T^*_{3}(\omega_{0}/2),
\end{eqnarray}
\begin{widetext}
\begin{eqnarray}
\label{12}
T_{n}(\omega)=
(E_{3^3D_{1}}-E_{2^3P_{J_{f}}})^{3/2}
(E_{3^3D_{1}}-E_{2^3S_{1}}-\omega)^{1/2}
\omega^{1/2}
\sum\limits_{J_{n}M_{J_n}}
\langle 2^3P_{J_{f}} | \textbf{e}_{3}^*\textbf{D} | 3^3D_{J_{n}} \rangle
\\\nonumber
\times
\sum\limits_{kJ_{k}M_{J_k}}
\left\lbrace
\frac{
	\langle 3^3D_{J_{n}} | \textbf{e}_{2}\textbf{D} | k^3P_{J_{k}}  \rangle
	\langle k^3P_{J_{k}}  | \textbf{e}_{1}\textbf{D}  | 2^3S_{1} \rangle
}
{
	E_{k^3P_{J_{k}}}-E_{2^3S_{1}}-\omega
}
+
\frac{
	\langle 3^3D_{J_{n}} | \textbf{e}_{1}\textbf{D}| k^3P_{J_{k}}  \rangle
	\langle k^3P_{J_{k}}  | \textbf{e}_{2}\textbf{D}| 2^3S_{1} \rangle
}
{
	E_{k^3P_{J_{k}}}-E_{3^3D_{J_{n}}}+\omega
}
\right\rbrace
.
\end{eqnarray}
\end{widetext}
Here $ \omega_{0}=E_{3^3D_1}-E_{2^3S_1} $ is the two-photon transition energy, $ \Gamma_{3^3D_{1}} $ is the natural width of $ 3^3D_{1} $ state, and $ f_{\mathrm{res}}^{3^3D_{1}} $, $ f_{\mathrm{nr}}^{3^3D_{1(2)}} $, $ f_{\mathrm{nr}}^{3^3D_{1(3)}} $ are coefficients that determine the dependence on angles $ \theta_{ij} $ ($ i,j=1,\;2,\;3 $), see Eqs. (13)-(17) and Appendix A in \cite{twophoton2021}. According to \cite{twophoton2021} the coefficient $ f_{\mathrm{res}}^{3^3D_{1}} $ describes the angular correlation of leading resonant two-photon scattering process via intermediate state with $ J_{n}=1 $, while $ f_{\mathrm{nr}}^{3^3D_{1(2)}} $ (or $ f_{\mathrm{nr}}^{3^3D_{1(3)}} $ ) corresponds to the interference contribution between fine sublevels with $ J_{n}=1 $ and $ J_{n}=2 $ (or $ J_{n}=1 $ and $ J_{n}=3 $). 

The first Lorentzian term in Eq. (\ref{11}) corresponds to the resonant transition, while two latter terms describe the asymmetry of line profile and were not considered in \cite{Dorrer, Hlousek}. Consequently, this leads to the shift of line position. Then the resonant transition frequency $ \omega_{\mathrm{res}} $ can be defined from $ d\sigma(\omega) $ in one of the evident ways as the maximum value of $ d\sigma(\omega) $ using the condition:
\begin{eqnarray}
\label{15}
\frac{d}{d\omega}\sigma(\omega)=0
.
\end{eqnarray}

In the resonant approximation, retaining only the first term in Eq. (\ref{11}), we find $\omega_{\mathrm{res}}=\omega_{\mathrm{max}}=\omega_{0}/2=(E_{3^3D_1}-E_{2^3S_1})/2 $. Then, taking into account the two interference terms in Eq. (\ref{11}) and solving Eq. (\ref{15}) with respect to $\omega$ one can find $ \omega_{\mathrm{max}} $ as
\begin{eqnarray}
\label{omegamax}
\omega_{\mathrm{max}}=(\omega_{0}-\delta_{\mathrm{NR}})/2
,
\end{eqnarray}
where
\begin{eqnarray}
\label{nr}
\delta_{\mathrm{NR}}=
\frac{f^{3^3D_{1(2)}}_{\mathrm{nr}}}{f^{3^3D_1}_{\mathrm{res}}}\frac{(\Gamma_{3^3D_1})^2}{4\Delta_{12}}
+
\frac{f^{3^3D_{1(3)}}_{\mathrm{nr}}}{f^{3^3D_1}_{\mathrm{res}}}\frac{(\Gamma_{3^3D_1})^2}{4\Delta_{13}}
.
\end{eqnarray} 
The nonresonant correction Eq. (\ref{nr}) is obtained as the contribution of the leading order in the series expansion in terms of the relation $ \Gamma/\Delta $ \cite{S-2020-importance}, when this parameter is small (in the considered examples inequality $\Gamma/\Delta\ll 1$ always holds). Performing angular algebra, we find 
\begin{eqnarray}
\label{f1}
f^{3^3D_{1(2)}}_{\mathrm{nr}}/f^{3^3D_1}_{\mathrm{res}}=95/198
,
\end{eqnarray}
and
\begin{eqnarray}
\label{f2}
f^{3^3D_{1(3)}}_{\mathrm{nr}}/f^{3^3D_1}_{\mathrm{res}}=7/198
. 
\end{eqnarray}

Equations (\ref{f1}) and (\ref{f2}) are calculated in a fully analytical way using definitions 
Eqs. (\ref{fres})-(\ref{12}). The summation over the entire spectrum of intermediate states is not required in the considered case since the corresponding contribution vanishes in the ratio $f_{\mathrm{nr}}/f_{\mathrm{res}}$. In the nonrelativistic limit sum over index $k$ in the curly brackets of Eq. (\ref{1}) (or Eq. (\ref{12})) becomes independent on the values of total angular momenta $J_{k},\;J_{n},\;J_{f}$.  Then after the reduction of matrix elements with the use of Eckart-Wigner theorem the corresponding reduced sums become equivalent for different values $J_{n}$ in Eq. (\ref{12}) and vanishes in the ratio $f_{\mathrm{nr}}/f_{\mathrm{res}}$. 

Finally, substituting the values for natural line width $ \Gamma_{3^3D_1}= 11.35(6)$ MHz \cite{Schmoranzer_1991} and energy differences $\Delta_{12} =1325.025(33) $ MHz and $ \Delta_{13} =  1400.290(33)$ MHz \cite{yerokhin2020} in Eq. (\ref{nr}), the NR correction to the transition frequency $2^3S_{1} -3^3D_{1}$ is
\begin{eqnarray}
\label{nr2}
\delta_{\mathrm{NR}}=0.0124(4)\,\,\, {\rm MHz}.
\end{eqnarray}

An analytical evaluation, Eqs. (\ref{f1}) and (\ref{f2}), shows that similarly to the NR correction to the $ 2s_{1/2}^{F=1}\rightarrow nd_{3/2(5/2)}^{F=2} $ transition frequencies in hydrogen \cite{solovyev2020proton}, the correction Eq. (\ref{nr}) also does not depend on the angles between any pair of the vectors $ \bm{\nu}_3 $, $\textbf{e}_1 $ and $ \textbf{e}_2 $. Therefore, the asymmetry of observed line profile cannot be eliminated by choosing 'magic' angles or experimental geometry as in \cite{beyer,yost}. An important outcome of these calculations, involving the natural level width, is that the magnitude of effect is at the level of experimental uncertainty $0.056$ MHz \cite{Dorrer}. 

However, the actual experimental width $ \Gamma^{\mathrm{exp}} $ of the observed profile differs significantly from the natural one $ \Gamma^{\mathrm{nat}} $ due to different broadening mechanisms \cite{riehle}. In fact, the width of level in Eq. (\ref{11}) must be associated with an experimental value \cite{deB-2,PRA-LSPS}. In \cite{Dorrer}, the main broadening effects are due to pressure and transit time. Denoting two latter contributions as  $ \Gamma^{\mathrm{pb}} $ and $ \Gamma^{\mathrm{tt}} $, respectively, the full width at half maximum in the experiment \cite{Dorrer} can be expressed as a sum of three contributions  
\begin{eqnarray}
\label{gamexp}
\Gamma^{\mathrm{exp}}=\Gamma^{\mathrm{nat}}+\Gamma^{\mathrm{pb}}+\Gamma^{\mathrm{tt}},
\end{eqnarray}
where $\Gamma^{\mathrm{nat}}$ denotes the natural level width.

According to \cite{Dorrer}, the pressure broadening is parametrized as $ \Gamma^{\mathrm{pb}}/p=35.7(1.7)\;[\mathrm{MHz/Torr}] $, where $ p $ is the pressure in Torr. Typically, the absorption signal is measured at various values of $ p $ and then the result is extrapolated to zero pressure (in \cite{Dorrer} the $p$ values are in the range of $ 0.05-0.5 $ Torr). 

The value of transit time broadening $ \Gamma^{\mathrm{tt}} $ is not presented in the experiment \cite{Dorrer}. However, we can roughly estimate $ \Gamma^{\mathrm{tt}} $ by attributing it to a difference between experimental width extrapolated to zero pressure $ \Gamma^{\mathrm{exp}}=11.33(19)$ MHz and natural width $ \Gamma^{\mathrm{nat}}=11.26$ MHz calculated theoretically \cite{drakebook}, that gives $ \Gamma^{\mathrm{tt}}=0.07(19) $ MHz. Thus, one can conclude that pressure broadening represents the main contribution to $ \Gamma^{\mathrm{exp}} $. Finally, for the pressures $p$ in the range from  $ 0.05$ to $0.5 $ Torr the experimental width Eq. (\ref{gamexp}) of observed profile belongs to the interval $ \Gamma^{\mathrm{exp}}\in [13.2(4),29.2(4)] $ MHz. Substitution of these values into Eq. (\ref{nr}) leads to the NR corrections $ \delta_{\mathrm{NR}}$ to transition frequency in the range from $0.016(1)$ to $0.082(19)$ MHz. The latter value only partly removes the current discrepancy between theoretical and experimental value of $2^3S_{1} -3^3D_{1}$ transition frequency which is of about $ 0.5 $ MHz \cite{yerokhin2021}.

The observed fluorescent signal was also fitted at a pressure of $ p=0.151 $ Torr (see Fig. 1 in \cite{Dorrer}). The NR correction corresponding to $ p=0.151 $ Torr is $ \delta_{\mathrm{NR}}=0.027 $ MHz. This value still does not remove the discrepancy between theoretical calculations and experiment found in \cite{yerokhin2021}, however reaches the level of experimental uncertainty $E^{\mathrm{exp}}(2^3S_{1}-3^3D_{1})=786\,823\,850.002(56) $ MHz \cite{Dorrer}. It is important to note that the resulting experimental value of $2^3S_1-3^3D_1$ transition frequency in \cite{Dorrer} was obtained by extrapolating the line position to zero pressure, i.e. to the case when $ \Gamma^{\mathrm{exp}} $ approximately equal $ \Gamma^{\mathrm{nat}} $. Then repairing the center of the line by NR correction at each value of pressure in accordance with Eq. (\ref{omegamax}), the result is expected to be different (see Fig. 2b in \cite{Dorrer}), eliminating the current discrepancy with theory, at least in part. We schematically reproduce fit given in \cite{Dorrer} and draw a new fit accounting for NR frequency corrected to $ \delta_{\mathrm{NR}}$ at each value of pressure according to $\omega'_{0}=\omega_{0}-\delta_{\mathrm{NR}} $, see Fig. 1 (blue points and corresponding line). The result of such extrapolation to zero pressure for shifted value $\omega'_{0}$ is, however, on the level of experimental uncertainty. Subtracting $\delta_{\mathrm{NR}}$ from experimental points $ \omega_{0} $ in Fig. 1 (blue points) we proceed from the fact that in experiment \cite{Dorrer} the nonresonant corrections were not considered, which means that they are included in the measured value (red points in Fig. 1).

\begin{figure}[hbtp]
  \caption{Extrapolations of line position for $2^3S_{1}-3^3D_{1}$ transition versus the pressure. The blue points take into account the corresponding NR corrections to the original data (red dots), given in accordance with Eq. (\ref{omegamax}). The corresponding dashed blue line represents the extrapolation to zero pressure through the points  that take into account the NR effects.  Likewise, the solid red line extrapolates uncorrected points. Uncertainties are also taken from \cite{Dorrer}. The frequency points take into account the second order Doppler effect, which turned out to be 8.1 KHz at T=300 K \cite{Dorrer}.  The least squares approach is used for extrapolation.}
  \centering
  \includegraphics[scale=0.85]{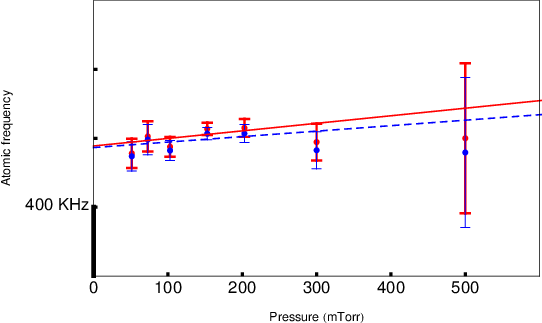}
\end{figure}

Similar calculations of NR corrections can be done for the measurement of $2^3S_1-4^3D_1$ and $2^3S_1-5^3D_1$ transition frequencies presented in \cite{Hlousek}. These experiments were carried out using the same technique as \cite{Dorrer} resulting to the values $ \Delta E^{\mathrm{exp}}(2^3S_{1}-4^3D_{1})=947\,000\,197.11(1.8)  $ MHz and $ \Delta E^{\mathrm{exp}}(2^3S_{1}-5^3D_{1})=102\,112\,869\,7.31(2.4)  $ MHz.  The pressure-broadening coefficients of $2^3S_1-4^3D_1$ and $2^3S_1-5^3D_1$ lines were determined as $ \Gamma^{\mathrm{pb}}/p=68.1(2.7)[\mathrm{MHz/Torr}] $ and $ \Gamma^{\mathrm{pb}}/p=78.5(2.7)[\mathrm{MHz/Torr}] $, respectively (see Table III in \cite{Hlousek}), while transit-time effects are supposed to be negligible. Then using the values of the corresponding fine structure intervals $ \Delta_{12}=E_{4^3D_{1}}-E_{4^3D_{2}}=555.231(7) $ MHz, $\Delta_{13}=E_{4^3D_{1}}-E_{4^3D_{3}}=591.253(6)$ MHz \cite{Morton}, and the natural level width $ \Gamma_{4^3D_{1}}^{\mathrm{nat}}=4.96274 $ MHz \cite{Theodosiou} we find the NR correction to the transition frequency $2^3S_1-4^3D_1$ in the range from $0.350(23)$ to $2.65(2)$ MHz for pressure $ p=0.5-1.5 $ Torr. The indicated corrections are at the level or exceed the experimental uncertainty for $ 2^3S_{1}-4^3D_{1} $ transition frequency which is $ 1.8 $ MHz. Likewise, the NR correction values for the $ 2^3S_{1}-5^3D_{1} $ transition frequency start at $0.79(5)$ and end at $6.6(4)$ MHz for the energy intervals $ \Delta_{12}=E_{5^3D_{1}}-E_{5^3D_{2}}=283.560(8)  $ MHz, $\Delta_{13}=E_{5^3D_{1}}-E_{5^3D_{3}}=302.781(8) $ MHz \cite{Morton}, and the natural level width $ \Gamma_{5^3D_{1}}^{\mathrm{nat}}=2.61381  $ MHz \cite{Theodosiou}. The results for all considered examples are summarized in Table \ref{tab1}. 
It is should be noted that considered NR corrections to $ 2^3S_{1}-n^3D_{1} $ ($ n=4,\;5 $) transition frequencies are higher than for $ 2^3S_{1}-3^3D_{1} $ transition frequency due to the smaller value of the corresponding fine structure intervals.
\begin{widetext}
\begin{center}
\begin{table}
\caption{Range of nonresonant corrections to $ 2^3S_{1}-n^3D_{1} $ ($ n=3,\;4,\;5 $) (5th column) transition frequencies (2nd column) at different ranges of experimental transition width (4th column), see Eq. (\ref{gamexp}). All values are in MHz. Uncertainties are given in brackets.}
\begin{tabular}{c c c c l}
\hline
\hline
Transition & Experiment, MHz \cite{Dorrer,Hlousek} & Theory, MHz \cite{yerokhin2020,yerokhin2021} & Experimental widths $ \Gamma^{\mathrm{exp}} $, MHz & NR corrections $ \delta_{\mathrm{NR}} $, MHz \\
\hline
$ 2^3S_{1}-3^3D_{1} $ & $ 786\,823\,850.002(56)    $ & $ 786\,823\,849.540(57) $ & $ 13.2(4)-29.2(4) $ & $ 0.016(1)-0.082(19) $\\
$ 2^3S_{1}-4^3D_{1} $ & $ 947\,000\,197.11(1.8)    $ & $947\,000\,194.44(5) $ & $ 39(1.4)-107(4)$ & $ 0.350(23)-2.65(2) $\\
$ 2^3S_{1}-5^3D_{1} $ & $ 102\,112\,869\,7.31(2.4) $ & $ 102\,112\,869\,8.36(5) $ & $ 41.9(1.4)-120(4)$ & $ 0.79(5)-6.6(4) $\\
\hline
\end{tabular}
\label{tab1}
\end{table}
\end{center}
\end{widetext}

Summing up the results of this work, it can be argued that the achieved measurement accuracy of the transition frequencies in helium has reached a level at which the nonresonant corrections and, in particular, the effect of quantum interference as their most significant part could be observed in two-photon atomic spectroscopy of helium. Since the NR correction found here is dependent on experimental conditions, each such calculation must take into account the corresponding features of experiment. The above analysis shows that the experimentally observed line broadening can play a key role in determining the nonresonant correction and, as a consequence, in determining the value of transition frequency. A decrease in the experimentally observed linewidth to the natural one should significantly reduce the risk of detecting discrepancies between the experimental and theoretical values of the transition energies.

Despite the fact that the experiments \cite{Dorrer, Hlousek} do not take into account the nonresonant effects, taking them into account still cannot completely remove the currently existing discrepancies with the theoretical predictions of the transition frequencies. Then it can be expected that the reconciliation between experiment and theory is most likely related to the issue of more accurate measurement or precise QED calculations of unaccounted contributions.

\section{Acknowledgements}
This work was supported by Russian Science Foundation (Grant No. 20-72-00003).

\bibliography{mybibfile}

\end{document}